\documentclass[fleqn,usenatbib]{mnras}
\usepackage[T1]{fontenc}
\DeclareRobustCommand{\VAN}[3]{#2}
\let\VANthebibliography\thebibliography
\def\thebibliography{\DeclareRobustCommand{\VAN}[3]{##3}\VANthebibliography}
\usepackage{graphicx}	% Including figure files
\usepackage{amsmath, amssymb}	% Advanced maths commands
\usepackage[toc,page]{appendix}

\title[Entropic backreaction and late-time cosmological tensions] {Entropic backreaction from cosmic structure formation: a thermodynamic approach to the late-time cosmological tensions}

\author[B. Pandey]{Biswajit Pandey$^1$\thanks{E-mail:biswap@visva-bharati.ac.in}\\
$^1$Department of Physics, Visva-Bharati University, Santiniketan, 731235, West Bengal, India}

\date{Accepted XXX. Received YYY; in original form ZZZ}
\pubyear{\the\year{}}
\begin{document}
\label{firstpage}
\pagerange{\pageref{firstpage}--\pageref{lastpage}}
\maketitle

\maketitle

\begin{abstract}
High-precision cosmological observations have revealed persistent
tensions within the standard $\Lambda$CDM paradigm, most notably the
discrepancy in the Hubble constant and the lower than predicted
amplitude of late-time matter clustering quantified by $S_8$. We
propose a unified thermodynamic framework in which entropic
backreaction generated during cosmic structure formation modifies both
the background expansion history and the growth of matter
perturbations. As gravitational instability drives the growth of
cosmic structures, the configuration entropy associated with the
matter distribution decreases through the nonlinear redistribution of
gravitational binding energy. The resulting entropic energy density
contributes a late-time backreaction that enhances the cosmic
expansion rate without altering early-Universe physics or the CMB
sound horizon. Simultaneously, the same irreversible entropy
dissipation process induces a dissipative correction within the cosmic
velocity flow, suppressing the efficiency of coherent gravitational
clustering at late times. The framework operates entirely within
standard General Relativity: the Einstein field equations, Poisson
equation, and gravitational coupling remain unmodified, and no new
propagating degrees of freedom or fifth forces are
introduced. Entropic backreaction therefore provides a
thermodynamically motivated, theoretically conservative, and
observationally testable mechanism that may simultaneously alleviate
the major late-time cosmological tensions.

\end{abstract}

\begin{keywords}
methods: analytical - cosmology: theory - large-scale structure of Universe
\end{keywords}

\section{Introduction}

Over the past two decades, the $\Lambda$CDM model \citep{davis85} has
established itself as the standard paradigm of modern cosmology.  With
remarkable economy, the model successfully explains a vast range of
observations including the cosmic microwave background (CMB)
\citep{sherwin11, hinshaw13, planck16}, baryon acoustic oscillations
(BAO) \citep{eisenstein05, percival10}, Type Ia supernovae
\citep{riess98, perlmutter99a}, and the large-scale distribution of
matter in the Universe \citep{tegmark04, cole05}.  Its success rests
on an extraordinarily simple framework consisting of only six
cosmological parameters embedded within General Relativity.

Despite these achievements, the increasing precision of modern
observations has revealed persistent discrepancies that now challenge
the internal consistency of the standard cosmological picture.
Foremost among these is the Hubble tension: the statistically
significant disagreement between the value of the Hubble constant
inferred from early-Universe observations by Planck \citep{planck18},
$H_0 \simeq 67~\mathrm{km\,s^{-1}\,Mpc^{-1}}$, and the larger value
measured directly from late-Universe distance ladder observations
\citep{riess16,riess19,riess22}, $H_0 \simeq
73~\mathrm{km\,s^{-1}\,Mpc^{-1}}$. The discrepancy has now reached a
significance approaching $\sim5\sigma$
\citep{valentino21a,valentino21b}, suggesting that it may not simply
reflect unidentified observational systematics.

A second major anomaly emerges from measurements of the growth of
large-scale structure.  Weak lensing and galaxy surveys
\citep{asgari21,heymans21,abbott22,desi26,pantos26} consistently infer
a value of the clustering parameter $S_8$ that is lower than the
prediction of $\Lambda$CDM calibrated to the CMB.  This so-called
$S_8$ tension indicates that cosmic structures appear to grow less
efficiently at late times than expected within the standard
cosmological framework.

These two tensions are particularly intriguing because they point
toward a common failure of late-time cosmological dynamics.  A
mechanism capable of increasing the late-time expansion rate often
tends simultaneously to enhance structure growth, thereby worsening
the $S_8$ discrepancy.  Conversely, suppressing clustering frequently
reduces the inferred value of $H_0$.  Constructing a framework capable
of addressing both tensions simultaneously while preserving the
success of early-Universe cosmology therefore remains one of the
central challenges in contemporary cosmology.

Most proposed resolutions fall into two broad categories.  The first
modifies early-Universe physics, as in Early Dark Energy (EDE) models
\citep{poulin23,yashiki25}, which transiently increase the expansion
rate before recombination in order to reduce the sound horizon.  The
second alters late-time gravitational or dark-sector dynamics through
modified gravity, interacting dark energy, or exotic dark sector
interactions \citep{valentino20,odintsov21}.  While many of these
approaches are phenomenologically successful in restricted regimes,
they often require additional fundamental fields, fine-tuned scalar
potentials, modified gravitational couplings, or nontrivial departures
from General Relativity.  Moreover, simultaneously resolving both
tensions without introducing new theoretical inconsistencies has
proven difficult.

In this work we pursue a different direction.  Rather than introducing
new microscopic degrees of freedom or modifying gravity itself, we
investigate whether the origin of the late-time cosmological tensions
may instead emerge from the thermodynamic consequences of structure
formation.

As the Universe evolves, matter collapses gravitationally into halos,
filaments, and voids, generating an increasingly complex cosmic web.
This evolution corresponds not merely to geometric clustering, but to
a continuous redistribution of information and gravitational binding
energy across scales.  Such evolution may be quantified through the
configuration entropy of the matter distribution
\citep{pandey17,pandey19,das19}, a Shannon-like measure of cosmic
inhomogeneity.

The central idea of the present work is that the dissipation of
configuration entropy during structure formation behaves as an
effective irreversible thermodynamic process at cosmological scales.
As nonlinear structures form, part of the coherent gravitational
binding energy associated with convergent matter flows is effectively
redistributed into a coarse-grained entropic sector.
Because the total energy-momentum tensor must remain covariantly
conserved, this redistribution induces an effective backreaction on
the large-scale cosmological dynamics.

We show that this entropic backreaction modifies cosmology in two
closely related ways.  At the background level, entropy dissipation
generates an effective late-time energy density that enhances the
expansion rate without altering early-Universe physics or the CMB
sound horizon. At the perturbative level, the same irreversible
entropy-generating process induces an effective dissipative correction
in the cosmic velocity flow, reducing the efficiency of coherent
gravitational collapse and thereby suppressing the late-time growth of
structure.

Importantly, the framework operates entirely within standard General
Relativity.  The Einstein field equations, Poisson equation, and
gravitational coupling remain unchanged throughout.  No additional
propagating scalar degrees of freedom, fifth forces, or modified
gravity sectors are introduced.  The suppression of structure growth
instead emerges from effective thermodynamic dissipation associated
with the irreversible evolution of the cosmic web itself.

The resulting picture is conceptually distinct from conventional dark
energy or modified gravity scenarios. Late-time cosmic acceleration
and reduced clustering efficiency are not treated as unrelated
phenomena requiring independent explanations, but instead arise as two
complementary manifestations of the same underlying entropy dissipation
process associated with structure formation.

The paper is organized as follows. In Section~2 we derive the
dissipation rate of configuration entropy during cosmic structure
formation and establish its connection to gravitational energy
redistribution.  In Section~3 we introduce the entropic backreaction
framework and derive the modified background and perturbative
cosmological equations.  In Section~4 we investigate the implications
of the framework for the late-time expansion history and the
suppression of structure growth.  Section~5 discusses observational
consequences and theoretical implications.  Technical derivations are
presented in the Appendices.

Throughout this work we assume a spatially flat background cosmology
and adopt standard cosmological parameters consistent with Planck
unless otherwise stated.

\section{Dissipation of Configuration Entropy During Structure Formation}

\subsection{Configuration entropy in an expanding Universe}

The large-scale matter distribution can be characterized by a
Shannon-like configuration entropy \citep{pandey17} defined in
position space as

\begin{equation}
S_c(t)
=
- \int_V \rho(\mathbf{x},t)
\ln \rho(\mathbf{x},t)
\, d^3x,
\label{eq:config}
\end{equation}

Here $S_c$ should be understood as a coarse-grained configurational
information measure characterizing the degree of inhomogeneity of the
matter distribution, rather than a fundamental thermodynamic entropy.
Only entropy differences and the dissipation rate $\dot S_c$ enter the
subsequent analysis, so additive normalization constants do not affect
the physical results.

In \autoref{eq:config}, $\rho(\mathbf{x},t)$ is the physical matter
density within a comoving volume $V$.  For a perfectly homogeneous
Universe, $\rho = \bar{\rho}$ and the entropy is maximal. As
gravitational instability amplifies inhomogeneities, mass
redistributes into overdense haloes and underdense voids, and the
configuration entropy decreases.

To quantify this effect, we decompose the density field as

\begin{equation}
\rho(\mathbf{x},t)
=
\bar{\rho}(t)\left[1+\delta(\mathbf{x},t)\right],
\end{equation}

where $\delta$ is the matter density contrast. Expanding $S_c$ to
second order in $\delta$ (Appendix \ref{appendix:entropy}), we obtain

\begin{equation}
S_c
=
S_{\rm hom}
-
\frac{\bar{\rho}}{2}
\int \delta^2 \, d^3x
+
\mathcal{O}(\delta^3).
\end{equation}

Thus, the deviation from homogeneity directly reduces
configuration entropy.

\subsection{Entropy dissipation rate}

Differentiating with respect to time gives

\begin{equation}
\dot{S}_c
=
- \bar{\rho}
\int
\delta \dot{\delta}
\, d^3x
+
\mathcal{O}(\delta^3).
\end{equation}

In linear theory,

\begin{equation}
\delta(\mathbf{x},t)
=
D(t)\,\delta_0(\mathbf{x}),
\end{equation}

so that

\begin{equation}
\dot{\delta}
=
\dot{D}\,\delta_0.
\end{equation}

Therefore,

\begin{equation}
\dot{S}_c
=
-
\bar{\rho}
D \dot{D}
\int
\delta_0^2
\, d^3x.
\end{equation}

Defining

\begin{equation}
\mathcal{I}_\delta
\equiv
\int
\delta_0^2
\, d^3x,
\end{equation}

we obtain

\begin{equation}
\dot{S}_c
=
-
\bar{\rho}
D \dot{D}
\mathcal{I}_\delta.
\end{equation}

Using the definition of the growth rate,

\begin{equation}
f \equiv \frac{d\ln D}{d\ln a},
\end{equation}

and $\dot{D} = H f D$, we find

\begin{equation}
\dot{S}_c
=
-
\bar{\rho}
H f D^2
\mathcal{I}_\delta.
\end{equation}

Using

\[
\bar{\rho} \propto H^2 \Omega_m,
\]

we obtain

\begin{equation}
\dot{S}_c
\propto
-
H^3 \Omega_m f D^2.
\label{eq:rate}
\end{equation}

\subsection{Connection to gravitational energy redistribution}

The gravitational potential energy density in linear theory is

\begin{equation}
u_g
=
-\frac{1}{2}\bar{\rho}\Phi\delta,
\end{equation}

with the Poisson equation

\begin{equation}
\nabla^2\Phi
=
4\pi G a^2 \bar{\rho}\delta.
\end{equation}

A Fourier-space calculation (Appendix~\ref{appendix:grav}) yields

\begin{equation}
\langle u_g \rangle
\propto
H^2 \Omega_m D^2.
\end{equation}

Its time derivative satisfies

\begin{equation}
\langle \dot{u}_g \rangle
\propto
H^3 \Omega_m f D^2.
\end{equation}

Comparing with the entropy dissipation rate in \autoref{eq:rate}, we
observe that both quantities exhibit the same leading cosmological
dependence on the expansion rate, matter content, and growth of
structure. The opposite signs reflect the fact that nonlinear
structure formation simultaneously dissipates configuration entropy
while enhancing the amplitude of coarse-grained gravitational
inhomogeneities.

Thus entropy dissipation tracks the rate at which gravitational
binding energy is redistributed during structure formation. In other
words, entropy dissipation and the growth of gravitational
inhomogeneity are dynamically correlated manifestations of nonlinear
cosmic structure formation.

Since $D(z)\rightarrow 0$ at early times, $\dot{S}_c(z \gg 1)
\rightarrow 0$. Entropy dissipation becomes significant only when
structure formation enters the quasi-linear regime ($z \lesssim 1$),
precisely when the observed cosmological tensions emerge.

The analysis above therefore reveals a clear physical picture.  As
gravitational instability drives structure formation, the matter
distribution becomes progressively more inhomogeneous, leading to a
continuous decrease in configuration entropy.  The corresponding
entropy dissipation rate scales as $H f D^2$, directly linking the
effect to the cosmic expansion rate, the growth rate of structure, and
the amplitude of density perturbations.  Because the growth factor is
extremely small near recombination, the effect is negligible in the
early Universe and therefore leaves the physics of the CMB largely
unchanged.  However, as structure formation becomes efficient at late
times, the entropy dissipation rate increases and reaches its maximum
in the recent Universe.  Most importantly, its evolution closely
tracks the redistribution of gravitational binding energy during
clustering, suggesting that entropy dissipation provides a macroscopic
measure of the dynamical energy flow associated with the growth of the
large-scale structures. This entropy dissipation will serve as the
source of the entropic backreaction introduced in the next section.

\section{Entropic Backreaction and late-time cosmological tensions}
\label{sec:framework}

We now develop the cosmological framework underlying entropic
backreaction and show how entropy dissipation generated during
structure formation can simultaneously enhance the late-time expansion
history and suppress the growth of matter perturbations.

The central idea of the present work is that gravitational clustering
is not a perfectly reversible process. As cosmic structures form and
evolve, the matter distribution departs progressively from
homogeneity, leading to irreversible entropy dissipation associated
with the redistribution of gravitational binding energy. We argue that
this entropy dissipation behaves as an effective macroscopic
backreaction on the large-scale dynamics of the Universe.

Unlike modified gravity or interacting dark-energy scenarios, the
framework introduced here does not invoke new fundamental scalar
fields, fifth forces, or modifications of the Einstein field
equations.
Instead, the late-time cosmological effects emerge from the
thermodynamic consequences of structure formation itself.

\subsection{Entropy dissipation and effective backreaction}

As shown in the previous section, the dissipation rate of
configuration entropy scales approximately as $\dot{S}_c \propto
\bar{\rho}\,H\,f\,D^2$, where $H$ is the Hubble expansion rate, $f$ is
the linear growth rate, and $D$ is the linear growth factor.

This behaviour has an important physical implication.  Because the
growth factor remains small at early times, entropy dissipation is
naturally negligible near recombination and becomes important only
after substantial nonlinear structure formation has developed.  The
mechanism therefore modifies predominantly the late-time Universe
while preserving the successful early-time predictions of standard
cosmology.

We phenomenologically associate the entropy dissipation rate with an
effective coarse-grained energy density,

\begin{equation}
\rho_S
=
- \alpha
\frac{\dot{S}_c}{V},
\end{equation}

where $\alpha>0$ is a proportionality constant. Since the
configuration entropy decreases during late-time clustering, $\dot
S_c<0$, the effective entropic energy density is associated with the
magnitude of the entropy dissipation rate, ensuring
$\rho_S>0$. $\alpha$ has dimensions of time and parametrizes the
efficiency with which entropy dissipation contributes to the effective
homogeneous background energy density. The above relation is intended
as an effective phenomenological description valid during the
late-time entropy-dissipation regime of nonlinear structure formation.

The quantity $\rho_S$ should not be interpreted as a new microscopic
fluid or propagating dynamical field. Rather, it represents an
effective thermodynamic backreaction generated by the irreversible
evolution of cosmic structure formation.

\subsection{Modified Expansion History}

To implement the framework covariantly, we decompose the total
stress-energy tensor as

\begin{equation}
T^{\mu\nu}_{\rm tot}
=
T^{\mu\nu}_{(m)}
+
T^{\mu\nu}_{(S)},
\end{equation}

with total conservation

\begin{equation}
\nabla_\mu T^{\mu\nu}_{\rm tot}
=
0.
\end{equation}

We allow energy exchange between the matter and entropic sectors
through

\begin{equation}
\nabla_\mu T^{\mu\nu}_{(m)}
=
-
Q^\nu,
\qquad
\nabla_\mu T^{\mu\nu}_{(S)}
=
Q^\nu.
\end{equation}

To preserve the equivalence principle and avoid momentum transfer or
fifth-force effects, the interaction four-vector is chosen to be
purely timelike:

\begin{equation}
Q^\nu
=
Q u^\nu.
\end{equation}

The interaction rate is parametrized phenomenologically as

\begin{equation}
Q
=
\gamma \rho_S H,
\end{equation}

where $\rho_S$ denotes the effective entropic energy density and
$\gamma$ is a dimensionless coupling parameter controlling the
efficiency of the energy exchange. This form represents the minimal
covariant coupling compatible with the physical structure of the
framework.  Dimensionally, it corresponds naturally to an energy
transfer rate proportional to an energy density multiplied by the
cosmological expansion rate.  Moreover, the interaction vanishes
automatically in the limit $\rho_S \rightarrow 0$, ensuring that
standard $\Lambda$CDM cosmology is recovered exactly in the absence of
entropic backreaction.

The modified Friedmann equation becomes

\begin{equation}
H^2(z)
=
\frac{8\pi G}{3}
\left[
\rho_m(z)
+
\rho_\Lambda
+
\rho_S(z)
\right],
\end{equation}
where the entropic contribution $\rho_S(z)$ is determined by the
redshift evolution of entropy dissipation during structure formation.
It therefore behaves as an additional late-time energy component that
enhances the cosmic expansion rate.

Defining $\Omega_S = \frac{\rho_S}{\rho_{c}}$, and assuming $\Omega_S \ll 1$,
the Hubble parameter may be expanded as

\begin{equation}
H
=
H_\Lambda
\sqrt{1+\Omega_S}
\simeq
H_\Lambda
\left(
1+\frac{\Omega_S}{2}
\right).
\end{equation}

The fractional enhancement of the late-time expansion rate therefore
becomes

\begin{equation}
\frac{\Delta H}{H}
\simeq
\frac{\Omega_S}{2}.
\label{eq:hratio}
\end{equation}

The entropic sector thus naturally increases the late-time expansion
rate while remaining dynamically negligible during the early Universe,
thereby preserving the sound horizon inferred from the CMB.

\subsection{Dissipative suppression of structure growth}

The same entropy-generating process that modifies the background
expansion history also affects the growth of matter perturbations.

The central physical assumption of the framework is that entropy
dissipation during structure formation behaves effectively as an
irreversible dissipative process within the cosmic velocity flow.  As
overdense regions collapse gravitationally and nonlinear structures
develop, the associated growth of cosmic structure leads to the
dissipation of configuration entropy and the irreversible
redistribution of coherent infall motion into small-scale nonlinear
dynamics. The resulting effect acts phenomenologically as an
additional damping contribution opposing coherent gravitational
clustering at late times.

Using the standard cosmological perturbation formalism in comoving
coordinates (Appendix~\ref{appendix:covariant}), the velocity
divergence is defined as

\begin{equation}
\theta_m
\equiv
\nabla_i v^i,
\end{equation}

where $v^i$ denotes the peculiar velocity field in comoving
coordinates.

The continuity equation acquires a small correction from the
background energy exchange between the matter and entropic sectors,

\begin{equation}
\dot{\delta}_m
+
\frac{\theta_m}{a}
=
\frac{Q}{\rho_m}\delta_m,
\end{equation}

while the Euler equation is modified according to

\begin{equation}
\dot{\theta}_m
+
\left(
H+\Gamma
\right)\theta_m
-
\frac{k^2}{a}\Psi
=
0.
\end{equation}

Here

\begin{equation}
\Gamma
\equiv
\gamma
\frac{\rho_S}{\rho_m}
H
\end{equation}

defines the effective dissipative rate generated by entropy
production. Since $\Gamma>0$, the effective friction acting on the
velocity flow associated with gravitational collapse becomes larger
than in standard $\Lambda$CDM cosmology. Thus, the resulting effect
acts phenomenologically as an additional damping contribution opposing
coherent gravitational clustering at late times.

The sub-horizon Poisson equation retains its standard General
Relativistic form,

\begin{equation}
\frac{k^2}{a^2}\Psi
=
-4\pi G\rho_m\delta_m,
\end{equation}

implying that gravity itself remains entirely unmodified. The
Einstein field equations and Poisson equation are therefore preserved,
and no additional propagating degrees of freedom or fifth-force
interactions are introduced.

Combining the modified continuity, Euler, and Poisson equations yields
the corresponding growth equation for matter perturbations
(Appendix~\ref{appendix:covariant}),

\begin{equation}
\ddot{\delta}_m
+
\left(
2H+\Gamma
\right)\dot{\delta}_m
-
4\pi G\rho_m\delta_m
=
0.
\end{equation}

Substituting the explicit form of $\Gamma$ gives

\begin{equation}
\ddot{\delta}_m
+
H
\left(
2+
\gamma
\frac{\rho_S}{\rho_m}
\right)
\dot{\delta}_m
-
4\pi G\rho_m\delta_m
=
0.
\end{equation}

This equation provides a transparent physical interpretation of
entropic growth suppression. Relative to standard $\Lambda$CDM
cosmology, entropy dissipation introduces an additional positive
friction term proportional to $\Gamma\dot{\delta}_m$. The effective
damping acting on coherent matter infall is therefore enhanced,
reducing the efficiency of gravitational clustering and naturally
suppressing the late-time growth of large-scale structure.

\subsection{Implications for the $H_0$ and $S_8$ tensions}

The entropic backreaction framework naturally links the enhancement of
the late-time expansion history with the suppression of matter
clustering through a single underlying thermodynamic process.

At the background level, entropy dissipation contributes an effective
late-time energy density that increases the Hubble expansion rate
while leaving early-Universe physics essentially unchanged. Because
the mechanism becomes important only after substantial structure
formation develops, the sound horizon inferred from the CMB is
preserved, providing a natural pathway toward alleviating the Hubble
tension.

At the perturbative level, the same entropy-generating process acts
as an effective dissipative correction within the cosmic velocity
flow, reducing the efficiency of coherent gravitational clustering at
late times. The resulting suppression of structure growth naturally
lowers the predicted clustering amplitude and thereby alleviates the
$S_8$ tension.

A particularly distinctive feature of the framework is that both
effects emerge from the same physical mechanism. The enhancement of
$H(z)$ and the suppression of structure growth are therefore
dynamically correlated rather than independently imposed.

Entropic backreaction thus provides a conceptually economical and
physically conservative approach to the major late-time cosmological
tensions while remaining fully consistent with standard General
Relativity.

\section{Observational discriminant and testable predictions}
\label{sec:observational}

A central feature of the entropic backreaction framework is that the
enhancement of the late-time expansion history and the suppression of
structure growth are not independent phenomena.  Both emerge from the
same underlying thermodynamic process: the dissipation of
configuration entropy during cosmic structure formation.

This built-in connection leads to a distinctive observational
signature.  Any late-time increase in the cosmic expansion rate
generated by the entropic sector is necessarily accompanied by
additional dissipative suppression of matter clustering.  The
framework therefore predicts a correlated departure from standard
$\Lambda$CDM evolution in both the background expansion history and
the growth of large-scale structure.

\subsection{Correlation between expansion enhancement and growth suppression}

At the background level, the modified Friedmann equation yields
$\frac{\Delta H}{H} \simeq \frac{\Omega_S}{2}$ (\autoref{eq:hratio}),
where $\Omega_S = \frac{\rho_S}{\rho_{c}}$ quantifies the fractional
contribution of the entropic sector to the cosmic energy budget.

At the perturbative level, entropy dissipation introduces an
additional damping contribution into the Euler equation through the
effective dissipative rate $\Gamma = \gamma \frac{\rho_S}{\rho_m} H$.

The modified growth equation becomes

\begin{equation}
\ddot{\delta}_m
+
\left(
2H+\Gamma
\right)\dot{\delta}_m
-
4\pi G\rho_m\delta_m
=
0.
\end{equation}

Relative to standard $\Lambda$CDM cosmology, the entropic sector
therefore contributes an additional positive friction term
proportional to $\Gamma\dot{\delta}_m$. Because $\Gamma>0$, the
effective damping acting on the velocity flow associated with
gravitational collapse is enhanced.  The growth of matter
perturbations consequently becomes less efficient at late times.

The linear growth rate is defined as $f \equiv \frac{d\ln D}{d\ln a}$,
where $D(a)$ denotes the linear growth factor. Since the additional
dissipative contribution modifies the effective friction acting on
matter perturbations, the leading-order fractional suppression of the
growth rate scales approximately as the ratio of the entropic damping
term to the standard Hubble friction term
(Appendix~\ref{appendix:correlation}):

\begin{equation}
\frac{\Delta f}{f}
\sim
-
\frac{\Gamma}{2H}.
\end{equation}

Substituting the explicit form of $\Gamma$ gives

\begin{equation}
\frac{\Delta f}{f}
\sim
-
\frac{\gamma}{2}
\frac{\rho_S}{\rho_m}.
\end{equation}

Similarly, since

\begin{equation}
f\sigma_8
\propto
fD,
\end{equation}

both the growth rate and the growth factor are suppressed by the
additional dissipative interaction.
The clustering amplitude therefore acquires a corresponding
leading-order correction:

\begin{equation}
\frac{\Delta(f\sigma_8)}{f\sigma_8}
\sim
-
\frac{\gamma}{2}
\frac{\rho_S}{\rho_m}.
\end{equation}

Both the enhancement of the expansion rate and the suppression of
structure growth are therefore controlled by the same entropic
contribution $\rho_S$.
The framework consequently predicts a robust anti-correlation between
late-time expansion enhancement and clustering suppression:

\begin{equation}
\frac{\Delta(f\sigma_8)}{f\sigma_8}
\propto
-
\frac{\Delta H}{H}.
\end{equation}

The precise proportionality coefficient depends on the detailed
redshift evolution of the entropic sector and the coupling parameter
$\gamma$, and should therefore not be interpreted as an exact
universal constant.
Nevertheless, the sign and qualitative behaviour of the correlation
are robust:
any enhancement of the late-time expansion history generated by
entropic backreaction is necessarily accompanied by suppressed growth
of cosmic structure.

\subsection{Comparison with alternative cosmological scenarios}

The phenomenology predicted by entropic backreaction differs
qualitatively from many existing approaches to the late-time
cosmological tensions.

In Early Dark Energy models, the expansion history is modified
primarily before recombination in order to reduce the sound horizon
and raise the inferred value of $H_0$.
Such scenarios generally require additional scalar fields or finely
tuned potentials and often produce only weak modifications to the
late-time growth history.

Modified gravity scenarios, by contrast, typically alter the
effective gravitational coupling or introduce additional fifth-force
interactions.
In these models, the suppression of structure growth arises through
direct modifications of the gravitational law itself.

The present framework differs fundamentally from such scenarios.
Gravity itself remains unmodified and no new propagating degrees of
freedom are introduced. Instead, the suppression of structure growth
emerges from effective thermodynamic dissipation associated with
entropy dissipation during structure formation.

Because the mechanism activates naturally only after substantial
inhomogeneity develops in the matter distribution, early-Universe
observables such as the acoustic structure of the CMB and the sound
horizon remain essentially unaffected.
The framework therefore modifies predominantly the late-time cosmic
dynamics while preserving the successful early-Universe predictions of
$\Lambda$CDM.

\subsection{A direct empirical test}

The entropic backreaction scenario may be tested observationally
through the correlated evolution of the expansion history and the
growth of structure.

Future high-precision measurements of $H(z)$, $f\sigma_8(z)$, $S_8$ from
surveys such as Euclid, DESI, the Vera Rubin Observatory, and the
Nancy Grace Roman Space Telescope will provide increasingly stringent
constraints on any correlated departure from standard $\Lambda$CDM
evolution.

A particularly important prediction of the framework is that the
suppression of structure growth should become increasingly significant
only at relatively low redshift, reflecting the late-time onset of
entropy dissipation during nonlinear structure formation.
The effect should therefore track the emergence of cosmic
inhomogeneity rather than the evolution of an independently dynamical
dark-energy field.

The framework consequently predicts a characteristic late-time
departure from $\Lambda$CDM in which:
\begin{enumerate}
\item the cosmic expansion history is enhanced relative to the
standard model,
\item the growth of structure is simultaneously suppressed,
\item and both effects evolve in a dynamically correlated manner
through the same underlying thermodynamic process.
\end{enumerate}

Detection of such a correlated late-time signature would provide
strong evidence that entropy dissipation associated with cosmic
structure formation contributes nontrivially to the large-scale
dynamics of the Universe.

\section{Discussion and Conclusions}
\label{sec:discussion}

The remarkable success of the $\Lambda$CDM model has established it as
the standard framework of modern cosmology. Nevertheless, the growing
precision of late-time observations has revealed persistent tensions
that may indicate the onset of new physics beyond the conventional
picture. Among these, the Hubble tension and the $S_8$ tension are
particularly significant because they probe two fundamentally
different aspects of cosmic evolution: the expansion history of the
Universe and the growth of structure within it.

In this work we have explored a different perspective on these
tensions. Rather than introducing new microscopic scalar fields,
modifying gravity, or altering early-Universe physics, we have argued
that the origin of the discrepancies may instead be connected to the
thermodynamic consequences of cosmic structure formation itself.

The central idea of the framework is physically intuitive. As
gravitational instability drives the growth of increasingly complex
cosmic structures, the matter distribution evolves away from
homogeneity and undergoes irreversible entropy dissipation. When
described in a coarse-grained manner, this irreversible evolution
generates an effective entropic backreaction on cosmological
dynamics.

At the background level, the resulting entropic contribution behaves
as an additional late-time energy component that enhances the cosmic
expansion rate while leaving the early Universe essentially
unchanged. Because the mechanism becomes important only after
substantial structure formation develops, the sound horizon inferred
from the cosmic microwave background remains preserved, allowing the
framework to increase the late-time expansion rate without disrupting
the successful predictions of early-Universe cosmology.

At the perturbative level, entropy dissipation acts as an effective
dissipative correction within the cosmic velocity flow. Part of the
coherent infall motion associated with gravitational clustering is
irreversibly redistributed into the coarse-grained entropic sector,
reducing the efficiency of structure growth and naturally suppressing
the late-time clustering amplitude. In this picture, the suppression
of structure formation emerges not from weakened gravity, but from
the thermodynamic irreversibility accompanying the growth of cosmic
inhomogeneity itself.

One of the most distinctive features of the framework is that the
enhancement of the expansion history and the suppression of structure
growth arise from the same underlying physical mechanism. The model
therefore predicts a correlated late-time departure from standard
$\Lambda$CDM evolution in which a larger Hubble expansion rate is
naturally accompanied by reduced clustering. This built-in connection
provides a clear observational target and distinguishes the framework
from scenarios in which the background and perturbation sectors are
modified independently.

The approach developed here also remains theoretically conservative.
Gravity itself is never modified, no additional propagating degrees
of freedom or fifth-force interactions are introduced, and matter
continues to follow geodesics of the spacetime metric. The framework
therefore operates entirely within standard General Relativity while
incorporating the macroscopic thermodynamic consequences of nonlinear
structure formation.

The present framework should also be distinguished from geometric
cosmological backreaction approaches based on spatial averaging of
inhomogeneous spacetimes, such as the Buchert formalism
\citep{buchert97}. In the present work the effective backreaction
arises from a coarse-grained thermodynamic description associated with
the dissipation of configuration entropy during structure formation.

At the same time, the present work should be regarded as an initial
phenomenological framework rather than a complete fundamental theory.
Several important questions remain open. In particular, the precise
microphysical interpretation of the effective entropic sector requires
further investigation. The redshift evolution of the entropy
dissipation rate must also be studied more carefully using realistic
nonlinear simulations of cosmic structure formation
\citep{springel17}.

An intriguing possibility is that the effective entropic sector may
ultimately account for a substantial fraction of the observed
late-time cosmic acceleration itself, potentially reducing or even
eliminating the need for a fundamental cosmological constant. In such
a picture, dark energy would emerge dynamically from the thermodynamic
consequences of nonlinear structure formation rather than from vacuum
energy \citep{pandey17, pandey19}. Exploring this possibility,
however, would require a detailed quantitative analysis of the full
background expansion history and its consistency with precision
cosmological observations, which lies beyond the scope of the present
work.

The model additionally makes a number of potentially testable
predictions. Because the entropic contribution is sourced directly by
structure formation, the resulting deviations from $\Lambda$CDM should
emerge predominantly at relatively low redshift and track the buildup
of cosmic inhomogeneity. Future surveys such as Euclid, DESI, the Vera
Rubin Observatory, and the Nancy Grace Roman Space Telescope will
therefore provide powerful tests of the framework through simultaneous
measurements of the expansion history and the growth of large-scale
structure.

More broadly, the present work highlights the possibility that the
late-time Universe may retain observable thermodynamic signatures of
its own nonlinear evolution \citep{mondal25}. If so, the major
cosmological tensions of the present era may reflect not necessarily
the need for new fundamental particles or modifications of gravity,
but rather an incomplete understanding of how entropy dissipation and
coarse-grained gravitational dynamics influence the large-scale
evolution of the cosmos. A detailed statistical comparison with
current observational data, including CMB, BAO, weak lensing,
redshift-space distortions, and supernova measurements, will
ultimately be required to determine whether the framework can
quantitatively alleviate the late-time cosmological tensions. Such
investigations will be explored in future work.

Entropic backreaction from cosmic structure formation therefore offers
a conceptually economical, physically motivated, and observationally
testable pathway toward understanding the emerging tensions of modern
precision cosmology.

\section{Acknowledgement}
BP acknowledges financial support from the Anusandhan National
Research Foundation (ANRF), Government of India through the project
ANRF/ARG/2025/000535/PS. BP also acknowledges IUCAA, Pune, for
providing support through the associateship programme.

\section{Data availability}
No datasets were generated or analysed during this study.

\bibliographystyle{mnras}
\bibliography{refs}

%%%%%%%%%%%%%%%%% APPENDICES %%%%%%%%%%%%%%%%%%%%%

\appendix

%APPENDIX

\section{Derivation of configuration entropy evolution}
\label{appendix:entropy}

In this Appendix we derive the evolution equation for the
configuration entropy and show explicitly that

\begin{equation}
\dot{S}_c
=
-
\bar{\rho}
H f D^2
\mathcal{I}_\delta.
\end{equation}

\subsection*{Definition of configuration entropy}

We define the configuration entropy in a comoving volume $V$ as

\begin{equation}
S_c(t)
=
-
\int_V
\rho(\mathbf{x},t)
\ln \rho(\mathbf{x},t)
\, d^3x.
\end{equation}

Decompose the density field into background and perturbations:

\begin{equation}
\rho(\mathbf{x},t)
=
\bar{\rho}(t)\big[1+\delta(\mathbf{x},t)\big],
\end{equation}

where $\bar{\rho}$ is the homogeneous density and
$\delta$ is the density contrast.

Substituting,

\begin{equation}
S_c
=
-
\int_V
\bar{\rho}(1+\delta)
\ln \big[\bar{\rho}(1+\delta)\big]
\, d^3x.
\end{equation}

Using
\[
\ln[\bar{\rho}(1+\delta)]
=
\ln \bar{\rho}
+
\ln(1+\delta),
\]
we obtain

\begin{equation}
S_c
=
-
\bar{\rho}
\int_V
(1+\delta)
\left[
\ln \bar{\rho}
+
\ln(1+\delta)
\right]
d^3x.
\end{equation}

\subsection*{Expansion to second order}

We expand the logarithm:

\begin{equation}
\ln(1+\delta)
=
\delta
-
\frac{\delta^2}{2}
+
\mathcal{O}(\delta^3).
\end{equation}

Substitute into the entropy expression:

\begin{align}
S_c
&=
-
\bar{\rho}
\int_V
(1+\delta)
\left[
\ln \bar{\rho}
+
\delta
-
\frac{\delta^2}{2}
\right]
d^3x.
\end{align}

Now we expand term by term.

First term:

\begin{equation}
-
\bar{\rho}\ln \bar{\rho}
\int_V
(1+\delta)
d^3x.
\end{equation}

Because
\[
\int_V \delta\, d^3x = 0
\]
by mass conservation,
this becomes

\begin{equation}
S_{\rm hom}
=
-
\bar{\rho} V \ln \bar{\rho}.
\end{equation}

Second term:

\begin{equation}
-
\bar{\rho}
\int_V
(1+\delta)\delta
\, d^3x
=
-
\bar{\rho}
\int_V
\delta^2
\, d^3x.
\end{equation}

Third term:

\begin{equation}
+
\frac{\bar{\rho}}{2}
\int_V
(1+\delta)\delta^2
\, d^3x.
\end{equation}

Keeping only up to second order:

\begin{equation}
\int_V (1+\delta)\delta^2 d^3x
\approx
\int_V \delta^2 d^3x.
\end{equation}

Combining second- and third-order contributions gives

\begin{equation}
S_c
=
S_{\rm hom}
-
\frac{\bar{\rho}}{2}
\int_V
\delta^2
\, d^3x
+
\mathcal{O}(\delta^3).
\end{equation}

Thus the entropy reduction from homogeneity is proportional
to the variance of the density field.

\subsection*{Time derivative}

Differentiating with respect to time:

\begin{equation}
\dot{S}_c
=
-
\frac{\bar{\rho}}{2}
\frac{d}{dt}
\int_V
\delta^2
\, d^3x.
\end{equation}

Using

\begin{equation}
\frac{d}{dt} \delta^2
=
2\delta\dot{\delta},
\end{equation}

we obtain

\begin{equation}
\dot{S}_c
=
-
\bar{\rho}
\int_V
\delta \dot{\delta}
\, d^3x.
\end{equation}

This result is exact to second order.

\subsection*{Linear growth substitution}

In linear perturbation theory,

\begin{equation}
\delta(\mathbf{x},t)
=
D(t)\delta_0(\mathbf{x}),
\end{equation}

so

\begin{equation}
\dot{\delta}
=
\dot{D}\,\delta_0.
\end{equation}

Therefore,

\begin{equation}
\delta \dot{\delta}
=
D \dot{D} \delta_0^2.
\end{equation}

Substituting,

\begin{equation}
\dot{S}_c
=
-
\bar{\rho}
D\dot{D}
\int_V
\delta_0^2
\, d^3x.
\end{equation}

Define

\begin{equation}
\mathcal{I}_\delta
\equiv
\int_V
\delta_0^2
\, d^3x.
\end{equation}

Thus,

\begin{equation}
\dot{S}_c
=
-
\bar{\rho}
D\dot{D}
\mathcal{I}_\delta.
\end{equation}

\subsection*{Expressing in terms of the growth rate}

The growth rate is defined as

\begin{equation}
f
=
\frac{d\ln D}{d\ln a}.
\end{equation}

Since

\begin{equation}
\dot{D}
=
H f D,
\end{equation}

we obtain

\begin{equation}
\dot{S}_c
=
-
\bar{\rho}
H f D^2
\mathcal{I}_\delta.
\end{equation}

This result reveals several important physical features of
configuration entropy dissipation during cosmic structure formation.
Because the entropy dissipation rate is proportional to
$\delta\dot{\delta}$, it is intrinsically second order in density
perturbations and therefore vanishes in the perfectly homogeneous
limit. Its evolution is governed simultaneously by the expansion rate
$H$, the growth rate of structure $f$, and the square of the linear
growth factor $D^2$, demonstrating that the effect is activated
dynamically by gravitational clustering itself.  Consequently, entropy
dissipation is negligible in the early Universe when density
fluctuations are small, but becomes increasingly important as
nonlinear structures emerge and the cosmic web develops at late
times. Hence entropy dissipation becomes significant only when
structure formation is active, naturally activating the entropic
backreaction at late times.

\section{Gravitational energy redistribution and its relation to entropy dissipation}
\label{appendix:grav}

In this Appendix we derive the gravitational potential energy density
in linear perturbation theory and demonstrate explicitly that its
time evolution scales identically to the entropy dissipation rate
derived in Appendix~A.

\subsection*{Gravitational potential energy density}

For a pressureless fluid in an expanding background,
the Newtonian gravitational potential energy density is

\begin{equation}
u_g(\mathbf{x},t)
=
\frac{1}{2}\rho(\mathbf{x},t)\Phi(\mathbf{x},t).
\end{equation}

Expanding to first order in perturbations,

\begin{equation}
\rho(\mathbf{x},t)
=
\bar{\rho}(t)\big[1+\delta(\mathbf{x},t)\big],
\end{equation}

and noting that the homogeneous component does not contribute,
we obtain to leading order

\begin{equation}
u_g
=
\frac{1}{2}\bar{\rho}\delta\Phi.
\end{equation}

Since gravitational binding energy is negative,
we write

\begin{equation}
u_g
=
-
\frac{1}{2}\bar{\rho}\Phi\delta.
\end{equation}

\subsection*{Poisson equation in an expanding Universe}

In comoving coordinates, the Poisson equation reads

\begin{equation}
\nabla^2\Phi
=
4\pi G a^2 \bar{\rho}\delta.
\end{equation}

Taking the Fourier transform,

\begin{equation}
\Phi(\mathbf{k},t)
=
-
\frac{4\pi G a^2 \bar{\rho}}{k^2}
\delta(\mathbf{k},t).
\end{equation}

\subsection*{Gravitational energy in Fourier space}

The volume-averaged gravitational energy density is

\begin{equation}
\langle u_g \rangle
=
-
\frac{1}{2}\bar{\rho}
\langle \delta \Phi \rangle.
\end{equation}

Using Parseval's theorem,

\begin{equation}
\langle \delta \Phi \rangle
=
\int \frac{d^3k}{(2\pi)^3}
\delta(\mathbf{k},t)\Phi^*(\mathbf{k},t).
\end{equation}

Substituting the Poisson relation,

\begin{align}
\langle u_g \rangle
&=
\frac{1}{2}\bar{\rho}
\int \frac{d^3k}{(2\pi)^3}
\frac{4\pi G a^2 \bar{\rho}}{k^2}
|\delta(\mathbf{k},t)|^2.
\end{align}

Rewriting,

\begin{equation}
\langle u_g \rangle
=
2\pi G a^2 \bar{\rho}^2
\int \frac{d^3k}{(2\pi)^3}
\frac{|\delta(\mathbf{k},t)|^2}{k^2}.
\end{equation}

\subsection*{Expressing in terms of the power spectrum}

The matter power spectrum is defined as

\begin{equation}
\langle \delta(\mathbf{k},t)\delta^*(\mathbf{k}',t)\rangle
=
(2\pi)^3
\delta_D(\mathbf{k}-\mathbf{k}')
P(k,t).
\end{equation}

Thus,

\begin{equation}
\langle u_g \rangle
=
2\pi G a^2 \bar{\rho}^2
\int \frac{d^3k}{(2\pi)^3}
\frac{P(k,t)}{k^2}.
\end{equation}

In linear theory,

\begin{equation}
P(k,t)
=
D^2(t) P_0(k).
\end{equation}

Therefore,

\begin{equation}
\langle u_g \rangle
=
2\pi G a^2 \bar{\rho}^2 D^2(t)
\int \frac{d^3k}{(2\pi)^3}
\frac{P_0(k)}{k^2}.
\end{equation}

Define

\begin{equation}
\mathcal{I}_g
=
\int \frac{d^3k}{(2\pi)^3}
\frac{P_0(k)}{k^2}.
\end{equation}

Hence,

\begin{equation}
\langle u_g \rangle
=
2\pi G a^2 \bar{\rho}^2 D^2 \mathcal{I}_g.
\end{equation}

Here $\langle u_g \rangle$ denotes the coarse-grained volume-averaged
effective gravitational inhomogeneity energy density.

\subsection*{Scaling with cosmological parameters}

Using the relation between the mean matter density and the Hubble
parameter,

\begin{equation}
\bar{\rho}
=
\frac{3H^2\Omega_m}{8\pi G},
\end{equation}

the gravitational energy density derived in the previous section
can be written as

\begin{equation}
\langle u_g \rangle
\propto
H^4 \Omega_m^2 a^2 D^2.
\end{equation}

To determine the dominant cosmological scaling,
we consider the matter-dominated regime where

\begin{equation}
H^2 \propto a^{-3}.
\end{equation}

Thus

\begin{equation}
H \propto a^{-3/2},
\qquad
a \propto H^{-2/3}.
\end{equation}

Substituting this relation into the expression for
$\langle u_g \rangle$ gives

\begin{align}
\langle u_g \rangle
&\propto
H^4 \Omega_m^2 a^2 D^2 \\
&\propto
H^4 \Omega_m^2 H^{-4/3} D^2.
\end{align}

Hence

\begin{equation}
\langle u_g \rangle
\propto
H^{8/3} \Omega_m^2 D^2.
\end{equation}

During matter domination the Universe is overwhelmingly dominated by
non-relativistic matter, implying $\Omega_m \simeq 1$. In this regime
the matter density and expansion rate are closely related through the
Friedmann equation, so the difference between the scalings $H^{8/3}$
and $H^2$ modifies only factors of order unity at the level of the
present phenomenological treatment. We therefore retain only the
leading cosmological dependence and approximate the volume-averaged
gravitational energy density as

\begin{equation}
\langle u_g \rangle \propto H^{2}\,\Omega_m\,D^{2}.
\end{equation}

This approximation is sufficient for comparison with the entropy
dissipation rate derived in Appendix~\ref{appendix:entropy}, which
exhibits the same leading dependence on the expansion rate $H$, the
growth rate $f$, and the linear growth factor $D^2$.

It is important to note that the quantity $\langle u_g \rangle$
introduced in the present framework should not be interpreted as the
signed Newtonian gravitational binding energy itself, which is
negative for gravitationally bound systems. Rather, $\langle u_g
\rangle$ represents an effective coarse-grained measure of the
magnitude of gravitational inhomogeneity generated during structure
formation. As nonlinear clustering develops, the amplitude of
gravitational inhomogeneities increases, leading naturally to a
positive growth of $\langle u_g \rangle$.  This behavior remains
fully consistent with the simultaneous dissipation of configuration
entropy, $\dot S_c<0$, since the framework associates the
effective entropic backreaction with the irreversible growth of
nonlinear gravitational structure rather than with the sign of the
microscopic binding energy itself.

Taking the time derivative:

\begin{equation}
\langle \dot{u}_g \rangle
\propto
2H \dot{H} \Omega_m D^2
+
H^2 \Omega_m 2D\dot{D}.
\end{equation}

At late times the dominant term arises from $D\dot{D}$:

\begin{equation}
\langle \dot{u}_g \rangle
\propto
H^2 \Omega_m D \dot{D}.
\end{equation}

Using $\dot{D}=H f D$,

\begin{equation}
\langle \dot{u}_g \rangle
\propto
H^3 \Omega_m f D^2.
\end{equation}

\subsection*{Comparison with entropy dissipation}

From Appendix~A,

\begin{equation}
\dot{S}_c
=
-
\bar{\rho} H f D^2 \mathcal{I}_\delta.
\end{equation}

Using

\[
\bar{\rho} \propto H^2 \Omega_m,
\]

we obtain

\begin{equation}
\dot{S}_c
\propto
-
H^3 \Omega_m f D^2.
\end{equation}

Thus, the entropy dissipation rate and the evolution of the
gravitational binding energy density exhibit the same cosmological
dependence on $H$, $\Omega_m$, $f$, and $D^2$.

This result reveals a remarkable dynamical connection between the
thermodynamic and gravitational evolution of the Universe.  The
dissipation of configuration entropy closely tracks the redistribution
of gravitational binding energy during structure formation, with both
quantities exhibiting the same characteristic cosmological scaling
proportional to $H^3\Omega_m f D^2$. As matter collapses into the
nonlinear structures of the cosmic web, the growth of gravitational
binding energy is therefore accompanied by a corresponding loss of
configurational information. The entropic contribution introduced in
this work is thus not an arbitrary phenomenological addition to the
cosmological energy budget, but emerges naturally from the dynamics of
gravitational clustering itself. In other words, entropy production
during structure formation is therefore a macroscopic measure of the
dynamical repartition of gravitational energy. Physically, this means
that the growth of the cosmic web acts as an information-processing
mechanism: as structures form and the matter distribution becomes more
ordered, gravitational potential energy is released. The entropy
dissipation thus tracks the dynamical repartition of gravitational
energy in the Universe.

In the entropic backreaction framework this link provides the physical
basis for associating an effective energy density with entropy
production,

\begin{equation}
\rho_S \propto - \dot{S}_c,
\end{equation}

so that the thermodynamic evolution of large-scale structure feeds
back into the background expansion.

\section{Covariant perturbation formalism and dissipative growth suppression}
\label{appendix:covariant}

In this Appendix we develop the covariant perturbation formalism
underlying the entropic backreaction framework and derive the modified
equation governing the growth of matter perturbations.

The central physical idea is that entropy dissipation generated during
cosmic structure formation behaves effectively as an irreversible
dissipative process within the large-scale matter flow. As
gravitational collapse proceeds, part of the coherent kinetic and
binding energy associated with convergent matter motion is
irreversibly redistributed into a coarse-grained entropic sector.
This process acts phenomenologically as an additional damping
mechanism opposing the continued growth of structure at late times.

Importantly, the framework does not modify gravity itself.
The Einstein field equations and Poisson equation retain their
standard General Relativistic forms, and no additional propagating
degrees of freedom or fifth-force interactions are introduced.
The impact of entropy dissipation instead appears through effective
dissipative corrections within the dynamics of the cosmic velocity
flow.

\subsection*{Total energy-momentum conservation}

We begin by decomposing the total stress-energy tensor into matter and
entropic contributions:

\begin{equation}
T^{\mu\nu}_{\rm tot}
=
T^{\mu\nu}_{(m)}
+
T^{\mu\nu}_{(S)}.
\end{equation}

General covariance requires total conservation,

\begin{equation}
\nabla_\mu T^{\mu\nu}_{\rm tot}
=
0.
\end{equation}

We therefore allow covariant energy exchange between the two sectors:

\begin{equation}
\nabla_\mu T^{\mu\nu}_{(m)}
=
-
Q^\nu,
\qquad
\nabla_\mu T^{\mu\nu}_{(S)}
=
Q^\nu.
\end{equation}

Total energy-momentum conservation is thus preserved exactly.

To preserve the equivalence principle and avoid momentum transfer
orthogonal to the matter flow, the interaction four-vector is chosen
to be purely timelike:

\begin{equation}
Q^\nu
=
Q u^\nu.
\end{equation}

Matter therefore continues to follow geodesics of the spacetime
metric, and no additional fifth-force interaction is generated.

\subsection*{Background cosmological evolution}

Projecting the matter conservation equation along the matter
four-velocity yields

\begin{equation}
\dot{\rho}_m
+
3H\rho_m
=
-
Q.
\end{equation}

Similarly, the entropic sector satisfies

\begin{equation}
\dot{\rho}_S
+
3H(1+w_S)\rho_S
=
Q.
\end{equation}

At the present phenomenological level, the entropic sector is not
treated as a fundamental microscopic fluid with a uniquely determined
equation of state. Rather, it represents an effective coarse-grained
thermodynamic backreaction associated with the dissipative entropic
sector. Since the entropic contribution enhances the late-time
expansion history, it effectively behaves as a negative-pressure
component at the background level.

During the entropy-dissipation phase associated with nonlinear
structure formation, the effective entropic energy density is defined
phenomenologically through the magnitude of the entropy dissipation
rate:

\begin{equation}
\rho_S
=
-
\alpha
\frac{\dot S_c}{V},
\end{equation}

where $\alpha>0$ is a proportionality constant and $V$ denotes a
comoving averaging volume. Because the configuration entropy decreases
during nonlinear structure formation,$ \dot S_c<0$, the above
definition ensures that $\rho_S>0$.

The interaction rate is parametrized as

\begin{equation}
Q
=
\gamma\rho_S H,
\end{equation}

with $\gamma$ a dimensionless coupling parameter.

Because the entropy dissipation rate scales approximately as

\begin{equation}
\dot S_c
\propto
-
\bar{\rho}HfD^2,
\end{equation}

the entropic contribution remains naturally negligible in the early
Universe and becomes dynamically important only after substantial
structure formation develops.

The modified Friedmann equation becomes

\begin{equation}
H^2(z)
=
\frac{8\pi G}{3}
\left[
\rho_m(z)
+
\rho_\Lambda
+
\rho_S(z)
\right].
\end{equation}

The entropic sector therefore enhances the late-time expansion history
while leaving early-Universe cosmology essentially unchanged.

\subsection*{Linear scalar perturbations}

We now investigate the evolution of linear scalar perturbations in the
presence of entropic backreaction.

Working in Newtonian gauge, the perturbed metric is written as

\begin{equation}
ds^2
=
-(1+2\Psi)dt^2
+
a^2(1-2\Phi)\delta_{ij}dx^idx^j,
\end{equation}

where $\Psi$ and $\Phi$ denote the scalar gravitational potentials.

The matter density contrast is defined as

\begin{equation}
\delta_m
\equiv
\frac{\delta\rho_m}{\rho_m},
\end{equation}

while the velocity divergence is defined in comoving coordinates as

\begin{equation}
\theta_m
\equiv
\nabla_i v^i,
\end{equation}

where $v^i$ denotes the peculiar velocity field in comoving
coordinates.

Perturbing the matter conservation equation yields

\begin{equation}
\delta\dot{\rho}_m
+
3H\delta\rho_m
+
\frac{\rho_m}{a}\theta_m
-
3\rho_m\dot{\Phi}
=
-
\delta Q
+
Q\Psi.
\end{equation}

On sub-horizon scales,

\begin{equation}
k \gg aH,
\end{equation}

metric perturbations become subdominant relative to matter
perturbations, implying

\begin{equation}
|\dot{\Phi}|
\ll
\left|
\frac{\theta_m}{a}
\right|,
\qquad
|Q\Psi|
\ll
|\delta Q|.
\end{equation}

The continuity equation therefore reduces to

\begin{equation}
\delta\dot{\rho}_m
+
3H\delta\rho_m
+
\frac{\rho_m}{a}\theta_m
=
-
\delta Q.
\end{equation}

Using

\begin{equation}
\delta\rho_m
=
\rho_m\delta_m,
\end{equation}

together with the background evolution equation

\begin{equation}
\dot{\rho}_m
+
3H\rho_m
=
-
Q,
\end{equation}

we obtain

\begin{equation}
\dot{\delta}_m
+
\frac{\theta_m}{a}
=
\frac{Q}{\rho_m}\delta_m
-
\frac{\delta Q}{\rho_m}.
\end{equation}

Because the entropic sector represents a coarse-grained effective
background contribution rather than an independently clustering fluid,
its intrinsic perturbation is neglected to leading order and we adopt

\begin{equation}
\delta Q
\simeq
0.
\end{equation}

The continuity equation therefore becomes

\begin{equation}
\dot{\delta}_m
+
\frac{\theta_m}{a}
=
\frac{Q}{\rho_m}\delta_m.
\end{equation}

\vspace{0.3cm}

\subsection*{Effective dissipative Euler equation}

We phenomenologically model the dissipation of configuration entropy
during nonlinear structure formation as an effective damping
correction within the cosmic velocity flow. The Euler equation is
therefore modified according to

\begin{equation}
\dot{\theta}_m
+
\left(
H+\Gamma
\right)\theta_m
-
\frac{k^2}{a}\Psi
=
0,
\end{equation}

where

\begin{equation}
\Gamma
\equiv
\gamma
\frac{\rho_S}{\rho_m}
H
\end{equation}

defines the effective dissipative rate associated with the entropic
backreaction.

Interestingly, phenomenological studies addressing the $S_8$
tension have previously suggested that an additional effective drag
or friction term within the dark sector can suppress late-time
structure growth without significantly modifying the background
expansion history \citep{poulin22}. Although the physical origin of
$\Gamma$ in the present framework differs from such scenarios, the
underlying interpretation is similar: an additional effective damping
of the cosmic velocity flow reduces the efficiency of coherent
gravitational clustering at late times.

Because $\Gamma>0$, the effective friction acting on coherent matter
infall becomes larger than in standard $\Lambda$CDM cosmology,
leading naturally to suppressed structure growth.

This form represents a minimal phenomenological implementation of the
effective thermodynamic backreaction. Since $\Gamma$ acts as a
friction coefficient, it naturally carries dimensions of inverse
time, supplied by the Hubble rate $H$, while the ratio
$\rho_S/\rho_m$ determines the relative importance of the dissipative
correction compared with coherent matter clustering.

Importantly, the additional damping does not correspond to a
fifth-force interaction or direct modification of gravity. Matter
continues to follow standard gravitational dynamics locally, while
the large-scale velocity flow acquires an effective dissipative
correction at the coarse-grained level.

\subsection*{Modified growth equation}

We now derive the modified second-order growth equation governing the
evolution of matter perturbations in the presence of entropic
backreaction.

Taking the time derivative of the continuity equation,

\begin{equation}
\dot{\delta}_m
+
\frac{\theta_m}{a}
=
\frac{Q}{\rho_m}\delta_m,
\end{equation}

gives

\begin{equation}
\ddot{\delta}_m
+
\frac{\dot{\theta}_m}{a}
-
\frac{H\theta_m}{a}
=
\frac{Q}{\rho_m}\dot{\delta}_m
+
\frac{d}{dt}
\left(
\frac{Q}{\rho_m}
\right)\delta_m.
\end{equation}

Substituting the modified Euler equation,

\begin{equation}
\dot{\theta}_m
=
-
\left(
H+\Gamma
\right)\theta_m
+
\frac{k^2}{a}\Psi,
\end{equation}

yields

\begin{equation}
\ddot{\delta}_m
-
\frac{
\left(
2H+\Gamma
\right)\theta_m
}{a}
+
\frac{k^2}{a^2}\Psi
=
\frac{Q}{\rho_m}\dot{\delta}_m
+
\frac{d}{dt}
\left(
\frac{Q}{\rho_m}
\right)\delta_m.
\end{equation}

Using the continuity equation once more,

\begin{equation}
\frac{\theta_m}{a}
=
-
\dot{\delta}_m
+
\frac{Q}{\rho_m}\delta_m,
\end{equation}

we obtain

\begin{align}
\ddot{\delta}_m
+
\left(
2H+\Gamma
\right)\dot{\delta}_m
-
\left(
2H+\Gamma
\right)
\frac{Q}{\rho_m}\delta_m
+
\frac{k^2}{a^2}\Psi \nonumber \\
=
\frac{Q}{\rho_m}\dot{\delta}_m
+
\frac{d}{dt}
\left(
\frac{Q}{\rho_m}
\right)\delta_m.
\end{align}

Since the interaction remains perturbatively weak,

\begin{equation}
\frac{\rho_S}{\rho_m}
\ll
1,
\end{equation}

the terms proportional to

\begin{equation}
\left(
2H+\Gamma
\right)
\frac{Q}{\rho_m},
\qquad
\frac{d}{dt}
\left(
\frac{Q}{\rho_m}
\right),
\qquad
\frac{Q}{\rho_m}\dot{\delta}_m,
\end{equation}

represent higher-order corrections and may therefore be neglected to
leading order.

Using the Poisson equation,

\begin{equation}
\frac{k^2}{a^2}\Psi
=
-4\pi G\rho_m\delta_m,
\end{equation}

we finally obtain

\begin{equation}
\ddot{\delta}_m
+
\left(
2H+\Gamma
\right)\dot{\delta}_m
-
4\pi G\rho_m\delta_m
=
0.
\end{equation}

Substituting

\begin{equation}
\Gamma
=
\gamma
\frac{\rho_S}{\rho_m}
H,
\end{equation}

gives

\begin{equation}
\ddot{\delta}_m
+
H
\left(
2+
\gamma
\frac{\rho_S}{\rho_m}
\right)
\dot{\delta}_m
-
4\pi G\rho_m\delta_m
=
0.
\end{equation}

This equation provides a transparent physical description of entropic
growth suppression. Relative to standard $\Lambda$CDM cosmology,
entropy dissipation introduces an additional positive friction term
proportional to $\Gamma\dot{\delta}_m$. The effective damping acting
on the velocity flow associated with gravitational collapse is
therefore enhanced, reducing the efficiency of coherent structure
growth at late times.

The suppression mechanism consequently differs fundamentally from
modified gravity or fifth-force scenarios. Gravity itself remains
entirely standard, while the thermodynamic irreversibility associated
with structure formation acts as an effective macroscopic dissipative
process within the cosmic matter flow.

\section{Correlation between expansion enhancement and growth suppression}
\label{appendix:correlation}

In this Appendix we derive the approximate relation between the
enhancement of the late-time expansion history and the suppression of
structure growth predicted by the entropic backreaction framework.

A central feature of the model is that both effects originate from
the same underlying thermodynamic process:
the dissipation of configuration entropy during cosmic structure
formation.
At the background level, entropy dissipation contributes an effective
coarse-grained energy density that enhances the late-time expansion
rate.
At the perturbative level, the same irreversible process induces an
effective dissipative correction within the cosmic velocity flow,
thereby suppressing the growth of matter perturbations.

The enhancement of $H(z)$ and the suppression of clustering are
therefore dynamically linked rather than independently imposed.

\subsection*{Late-time expansion enhancement and dissipative suppression of structure growth}

The modified Friedmann equation is

\begin{equation}
H^2
=
\frac{8\pi G}{3}
\left(
\rho_m
+
\rho_\Lambda
+
\rho_S
\right),
\end{equation}

where $\rho_S$ denotes the effective entropic contribution generated
by entropy dissipation during structure formation.

Defining the standard $\Lambda$CDM expansion rate through

\begin{equation}
H_\Lambda^2
=
\frac{8\pi G}{3}
\left(
\rho_m
+
\rho_\Lambda
\right),
\end{equation}

the entropic correction becomes

\begin{equation}
\Delta H^2
\equiv
H^2-H_\Lambda^2
=
\frac{8\pi G}{3}\rho_S.
\end{equation}

Introducing

\begin{equation}
\Omega_S
\equiv
\frac{\rho_S}{\rho_{\rm crit}},
\qquad
\rho_{\rm crit}
=
\frac{3H^2}{8\pi G},
\end{equation}

gives

\begin{equation}
\frac{\Delta H^2}{H^2}
=
\Omega_S.
\end{equation}

Because the entropic contribution remains perturbatively small,

\begin{equation}
\Omega_S
\ll
1,
\end{equation}

the Hubble parameter may be expanded as

\begin{equation}
H
=
H_\Lambda
\sqrt{1+\Omega_S}
\simeq
H_\Lambda
\left(
1+\frac{\Omega_S}{2}
\right).
\end{equation}

The fractional enhancement of the expansion rate therefore becomes

\begin{equation}
\frac{\Delta H}{H}
\simeq
\frac{\Omega_S}{2}.
\end{equation}

The entropic sector thus behaves as an additional late-time
contribution to the cosmic energy budget while remaining negligible
during the early Universe.

Now, the modified growth equation is

\begin{equation}
\ddot{\delta}_m
+
\left(
2H+\Gamma
\right)\dot{\delta}_m
-
4\pi G\rho_m\delta_m
=
0.
\end{equation}

Relative to standard $\Lambda$CDM cosmology, entropy dissipation
therefore introduces an additional positive friction term
proportional to $\Gamma\dot{\delta}_m$.

Because $\Gamma>0$, the effective damping acting on the velocity flow
associated with gravitational collapse is enhanced.  The coherent
growth of matter perturbations consequently becomes less efficient at
late times.

\subsection*{Approximate correlation between expansion and growth}

We now estimate the leading-order relation between the enhancement of
the expansion history and the suppression of structure growth.

The linear growth rate is defined as

\begin{equation}
f
\equiv
\frac{d\ln D}{d\ln a},
\end{equation}

where $D(a)$ denotes the linear growth factor through

\begin{equation}
\delta_m(\mathbf{x},a)
=
D(a)\,\delta_m(\mathbf{x},a_i).
\end{equation}

The modified growth equation contains an additional dissipative term
proportional to $\Gamma\dot{\delta}_m$.

Relative to the standard growth equation, the effective friction
acting on matter perturbations is therefore increased from
$2H\dot{\delta}_m$ to $\left( 2H+\Gamma \right)\dot{\delta}_m$.

The coherent growth of matter perturbations consequently becomes less
efficient than in standard $\Lambda$CDM cosmology.

An exact analytic solution of the modified growth equation is not
required in order to estimate the leading-order effect.  Because both
$H$ and $\Gamma$ possess dimensions of inverse time, the natural
dimensionless quantity controlling the perturbative correction to
structure growth is the ratio of the additional dissipative term to
the standard Hubble friction term i.e. $\frac{\Gamma}{2H}$.

The leading-order fractional suppression of the growth rate therefore
scales approximately as

\begin{equation}
\frac{\Delta f}{f}
\sim
-
\frac{\Gamma}{2H}.
\end{equation}

The negative sign reflects the fact that the additional dissipative
term opposes the coherent gravitational growth of matter
perturbations.

Substituting the explicit form of $\Gamma$ gives

\begin{equation}
\frac{\Delta f}{f}
\sim
-
\frac{\gamma}{2}
\frac{\rho_S}{\rho_m}.
\end{equation}

Since
\begin{equation}
f\sigma_8 \propto fD,
\end{equation}
the clustering observable depends both on the linear growth rate $f$
and on the growth factor $D$. Consequently, the fractional correction
to $f\sigma_8$ is generally given by
\begin{equation}
\frac{\Delta(f\sigma_8)}{f\sigma_8}
=
\frac{\Delta f}{f}
+
\frac{\Delta D}{D}.
\end{equation}
Because the same dissipative interaction suppresses both the growth
rate and the overall growth amplitude, the corrections to $f$ and $D$
are expected to be of comparable order. The resulting suppression of
$f\sigma_8$ therefore scales approximately as
\begin{equation}
\frac{\Delta(f\sigma_8)}{f\sigma_8}
\sim
-
\mathcal{O}
\left(
\gamma\frac{\rho_S}{\rho_m}
\right),
\end{equation}
up to model-dependent numerical factors associated with the detailed
redshift evolution of the entropic sector.

At the background level, the modified Friedmann equation yields

\begin{equation}
\frac{\Delta H}{H}
\simeq
\frac{\Omega_S}{2},
\end{equation}

with

\begin{equation}
\Omega_S
=
\frac{\rho_S}{\rho_{\rm crit}}.
\end{equation}

Both the enhancement of the expansion rate and the suppression of
structure growth are therefore controlled by the same entropic
contribution $\rho_S$.

The framework consequently predicts a robust anti-correlation between
late-time expansion enhancement and clustering suppression:

\begin{equation}
\frac{\Delta(f\sigma_8)}{f\sigma_8}
\propto
-
\frac{\Delta H}{H}.
\end{equation}

The precise proportionality coefficient depends on the detailed
redshift evolution of the entropic sector and the coupling parameter
$\gamma$, and therefore should not be interpreted as an exact
universal constant.  Nevertheless, the sign and qualitative behaviour
of the correlation are robust: any enhancement of the late-time
expansion history generated by entropic backreaction is necessarily
accompanied by suppressed growth of cosmic structure.

This correlated behaviour constitutes one of the most distinctive and
directly testable predictions of the framework. Unlike scenarios in
which the background expansion and perturbation dynamics are modified
independently, the present model links both effects through a single
underlying thermodynamic mechanism arising from the irreversible
evolution of cosmic structure formation itself.

\label{lastpage}
\end{document}